\documentclass[aps,prf,superscriptaddress,preprint, amsmath,amssymb]{revtex4-2}

\usepackage{graphicx}
\usepackage{dcolumn}
\usepackage{bm}
\usepackage{color}
\usepackage{hyperref}
\usepackage{amsmath}
\usepackage{empheq}

\begin{document}
\setlength{\fboxrule}{1.5pt}
\title{Chiral swimmer with a regular arbitrary active patch} 

\author{Shiba Biswas}
\affiliation{Department of Mathematics, Indian Institute of Technology Kharagpur, Kharagpur 721302, India}

\author{P. S. Burada}
\affiliation{Department of Physics, Indian Institute of Technology Kharagpur, Kharagpur 721302, India}

\author{G. P. Raja Sekhar}\email{rajas@iitkgp.ac.in}
\affiliation{Department of Mathematics, Indian Institute of Technology Kharagpur, Kharagpur 721302, India}

\begin{abstract}
We investigate the low Reynolds number hydrodynamics of a spherical swimmer with a predominantly hydrophobic surface, except for a hydrophilic active patch. This active patch covers a portion of the surface and exhibits chiral activity that varies as a function of $\theta$ and $\phi$. Our study considers two types of active patches: (i) a symmetric active patch (independent of $\phi$) and (ii) an arbitrary active patch (depends on both $\theta$ and $\phi$). The swimming velocity, rotation rate, and flow field of the swimmer are calculated analytically. The objective of this work is to find the optimal configurations for both patch models to maximize the swimmer's velocity and efficiency. Interestingly, the maximum velocity can be controlled by adjusting the hydrophobicity, patch configuration, and strength of the surface activity. We find that for the symmetric patch model, the swimmer's velocity is $U_{SP} = 1.414 U_s$, where $U_s$ is the velocity of a swimmer whose surface is fully covered with chiral activity as a reference. For the arbitrary patch model, the velocity is $U_{AP} = 1.45 U_s$, which is higher than that of the symmetric patch model. Our results indicate that swimmers with low hydrophobicity exhibit efficient swimming characteristics. Additionally, due to the incomplete coverage of the active patch, the Stokeslet and Rotlet terms appear in the flow field generated by the swimmer, which is a deviation compared to the case of a swimmer whose surface is fully covered with chiral activity. This study provides insights useful for designing synthetic active particles, which can be applied, for example, in targeted drug delivery, chemotaxis, and phototaxis.
\end{abstract}

\maketitle

\section{Introduction}
\label{sec: introduction}

The study of active swimmers is an emerging field of interest \cite{lauga2009hydrodynamics,vicsek2012collective,marchetti2013hydrodynamics, Eric_book}. 
A significant portion of these studies is predicated on the notion that each particle moves with a velocity determined by its activity and the ambient fluid flow. 
Various models have been proposed to study the self-propelled motion of diverse microorganisms \cite{Eric_book}.
Many theoretical, experimental, and numerical studies concentrated on shapes, surface properties, hydrodynamic interaction, and collective behaviors of microorganisms~\citep{stone1996propulsion,pedley2016spherical,theers2016modeling,liu2022migration}.
Microorganisms that swim inhabit an environment where viscous forces govern their movements. 
Lighthill \cite{lighthill1952squirming} and Blake \cite{blake1971spherical} first proposed the squirmer model to study the propulsion of ciliated microorganisms. When the surface of a motile microorganism is densely covered with beating cilia, e.g., {\it{Paramecium}} and {\it{Opalina}}, it generates metachronal waves, which cause an active slip velocity over the surface of the microorganism \cite{blake1971spherical,machemer1972ciliary}. 
Apart from this, there are other ciliated microorganisms, such as {\it{Marine Zooplankton}} and {\it{Acineta Protozoa}}, which have cilia at a particular region on their surface \cite{gibbons1981cilia}. 
{{For such complex distributions of cilia over the surface, microorganisms encounter abrupt changes in hydrodynamic slip and active slip between the ciliated region and the rest of the surface. Willmott \cite{willmott2008dynamics,willmott2009slip} studied the slip-induced dynamics of a Janus-like sphere in Stokes flow, focusing on binary discontinuous boundary conditions between slip and no-slip regions. The resulting flow asymmetry due to these discontinuous boundary conditions was analyzed. Dhar et al. \cite{dhar2019self} later examined the hydrodynamics of a rigid slip-stick swimmer at low Reynolds number, with active slip on an axisymmetric patch and Navier slip on the rest of the surface. They found that adjusting the patch configuration could enhance the swimmer's migration velocity by up to 50\% compared to a fully covered swimmer. In our recent work \cite{biswas2023hydrodynamics}, we explored a spherical swimmer in arbitrary Stokes flow with a non-axisymmetric patch dividing the surface into distinct slip regions. We showed that varying the patch configuration and slip length ratio effectively controls locomotion. The resulting jump in the boundary condition creates localized effects near the swimmer, influencing streamline deviations. Both studies identify optimal patch configurations that maximize swimming velocity. More recently, Yang et al. \cite{yang2023stick} studied a spherical swimmer with hydrophobic and hydrophilic regions and axisymmetric surface activity. They demonstrated that, by selecting appropriate swimming modes and cap shape, a swimmer with a hydrophobic cap can outperform a fully hydrophilic swimmer in efficiency. These investigations provide insight into the dynamics of active particles with regionally confined surface activity.}}

To our knowledge, there is no experimental study on the fabrication technique to implement artificial cilia over a portion of the surface of solid particles. However, few studies are based on the simulation of multiple cilia concentrated over the spherical surface to investigate microbial locomotion \cite{omori2020swimming,rode2021multi,westwood2021coordinated}. Apart from this, there are studies on the locomotion of self-propelled droplets fully or partially covered with a stagnant surfactant layer \cite{sadhal1983stokes,pak2014viscous,seemann2016self,sharanya2021transient}. There are chemically active microswimmers such as the electrophoretic colloidal particles (\cite{anderson1989colloid}), dielectric spheres partially covered by Pt \cite{howse2007self,ebbens2012size,brown2014ionic} or the Au-Pt Janus particles \cite{lee2014self}. These swimmers move mostly because of different phoretic effects \cite{moran2017phoretic}. The phoretic motion of a spherical particle where surface reactants are distributed asymmetrically is studied by Golestanian et al. \cite{golestanian2005propulsion}. Consecutively, by taking into account the activity and mobility, the phoretic motion of the colloidal particle has been studied when these are distributed symmetrically \cite{golestanian2007designing} and asymmetrically \cite{lisicki2018autophoretic} over the particle. {Within the framework of self-diffusiophoresis, Popescu et al. \cite{popescu2018effective} discussed several types of induced surface active slip velocity over the particle which mimic puller, pusher, and neutral swimmer. In this study, one of the considered spherical particle's surface is partitioned into two different regions by a chemically active cap, and the other region is inert. For this, the binary-valued phoretic mobility has been incorporated. As a result, discontinuity in the phoretic surface active slip arises. The flow field behavior in the laboratory frame of reference due to the chemically active cap has been analyzed.} Recently, Burada et al. \cite{burada_pre} studied the chiral swimmer where surface active slip velocity is distributed asymmetrically, contributing to both translations and rotations of the motion. This analogy between the structure of the swimmer's slip and the surface active slip velocity of a model artificial swimmer supports the motivation of the present framework.

Based on the literature and observations, we aim to develop a well-controllable spherical chiral swimmer driven by surface active slip concentrated over an arbitrary patch of its surface, with the rest being hydrophobic, characterized by Navier slip length. This study's primary objective is to identify the optimal configuration of the active patch region to maximize the swimmer's velocity strength. The article is organized as follows. 
In Section~\ref{sec: math_model}, we present our model system, including the relevant boundary conditions. Section~\ref{sec: Ana_solution} outlines our solution methodology. Section~\ref{sec: axi_model} focuses on a swimmer with a symmetric active patch. In Section~\ref{sec: Naxi_model}, we explore the case of a swimmer with an arbitrary active patch and determine the global optimal patch configuration that maximizes the migration velocity of the swimmer. Finally, the discussion and conclusions are presented in Sections~\ref{sec: discussion} and~\ref{sec: conclusions}, respectively.


\section{Model}
\label{sec: math_model}

\begin{figure}
    \centering
    \includegraphics[width=0.8\linewidth]{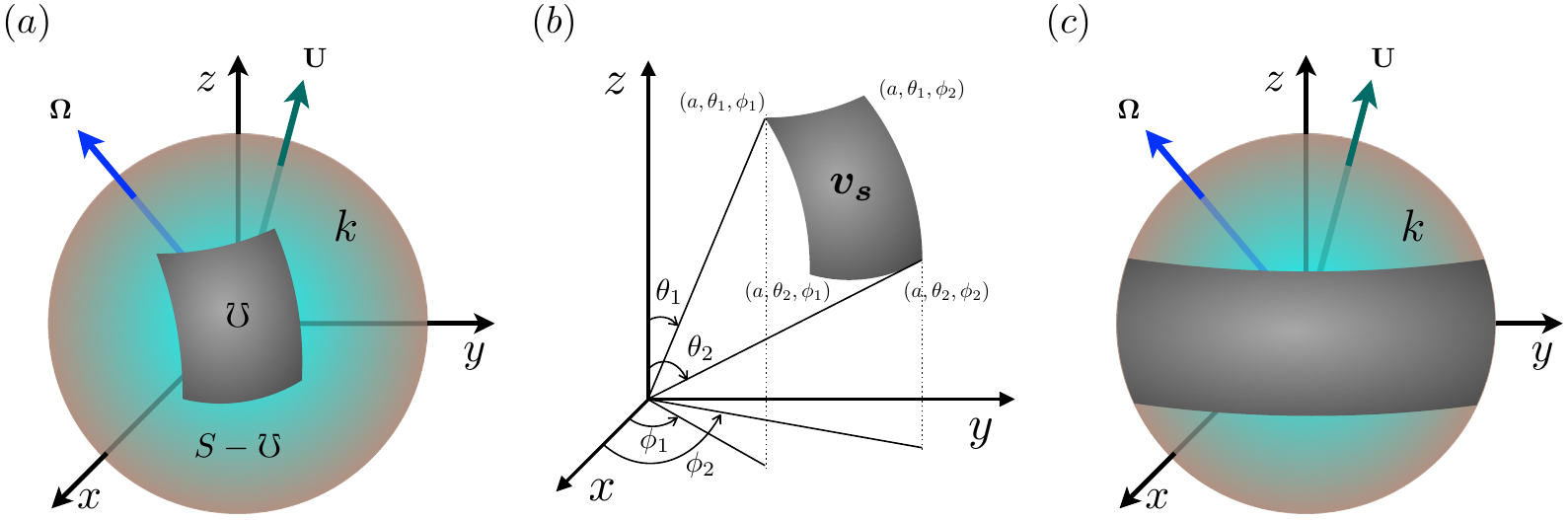}
    \caption{(a) Schematic diagram of a spherical swimmer with an arbitrary patch in the region $\mho$, where an active slip $\boldsymbol{v_s}$ is prescribed (b). In the other region $(S - \mho)$, the swimmer surface is hydrophobic with a slip length of $k$. (c) Swimmer with a symmetric patch.}
    \label{fig:  schematic}
\end{figure}
We consider a spherical swimmer of radius $a$ in a quiescent ambient viscous fluid with viscosity $\mu$. We assume that the activity on the surface is restricted to a patch region (refer to figure~\ref{fig:  schematic}(a,b)). In figure~\ref{fig:  schematic}(c), we show the corresponding model with a surface activity prescribed on a strip as a symmetric patch. We denote the patch region as $\mho$ and the rest of the surface of the swimmer as $S-\mho$, where $S$ denotes the surface of the spherical swimmer. We further assume that the region $S-\mho$ is hydrophobic with a Navier slip length $k \ll a$. {Note that $k=0$ corresponds to a smooth, hydrophilic surface, whereas $k \neq 0$ indicates a rough, hydrophobic surface~\cite{schaffel2016local,zhang2022surface}.} It is assumed that the Stokes equation describes the flow generated by the swimmer in the low Reynolds number regime in an incompressible Newtonian fluid as~\cite{lighthill1952squirming,blake1971spherical}
\begin{equation}
\mu{\boldsymbol{\nabla}}^{2}\boldsymbol{u} = \boldsymbol{\nabla}{p}, \quad {\boldsymbol{\nabla}}\cdot\boldsymbol{u} = 0\,,
\label{eqn: stokes}
\end{equation}
where $\boldsymbol{u}$ and $p$ are the corresponding velocity and pressure fields, respectively, and $\boldsymbol{\nabla}$ denotes the gradient operator. We use the spherical polar coordinates $(r,\theta,\phi)$ coordinates with the corresponding unit vectors $(\boldsymbol{e}_r,\boldsymbol{e}_\theta,\boldsymbol{e}_\phi)$. The active patch region is defined as $\mho = \big\{ (r,\theta,\phi) \in S~|~r=a,\theta_{1} \leqslant \theta \leqslant \theta_{2}, \phi_1 \leqslant \phi \leqslant \phi_2 \big\}$, where $\theta_{1},\theta_{2}$ are polar and $\phi_{1},\phi_{2}$ are azimuthal slip angles (refer to figure~\ref{fig:  schematic}(b)). The active slip $\boldsymbol{v_s}$ on the surface of the swimmer is prescribed, in the body frame of reference $(x,y,z)$, in spherical coordinates as~\cite{burada_pre}
\begin{eqnarray}
\boldsymbol{v_s}(\theta,\phi) = \sum_{n=1}^{\infty} \Big[
\beta_n \boldsymbol{\nabla}_s \tilde{S}_n (\theta,\phi) 
+ \gamma_n \boldsymbol{e}_r \times \boldsymbol{\nabla}_s \tilde{H}_n (\theta,\phi)\Big],
\label{eqn: active_slip}
\end{eqnarray}
where $\boldsymbol{\nabla}_s = \boldsymbol{e}_\theta {\partial}/{\partial \theta} + \boldsymbol{e}_ \phi(1/\sin \theta){\partial}/{\partial \phi}$ is the surface gradient operator, 
$\tilde{S}_n(\theta,\phi) = \sum^n_{m=0} \big(\tilde{A}_{nm} \cos(m\phi)+\tilde{B}_{nm}\sin(m\phi)\big) P^m_n(\cos \theta)$ and $\tilde{H}_n(\theta,\phi) = \sum^n_{m=0} \big(\tilde{C}_{nm}\cos(m\phi)+\tilde{D}_{nm}\sin(m\phi)\big) P^m_n (\cos \theta)$, are spherical harmonics. 
{The coefficients $\beta^A_{nm}=\beta_n \tilde{A}_{nm}, \beta^B_{nm}=\beta_n \tilde{B}_{nm}, \gamma^C_{nm}=\gamma_n \tilde{C}_{nm},\gamma^D_{nm}=\gamma_n \tilde{D}_{nm}$ are the mode amplitudes of the active slip, where $(\beta^A_{nm},\beta^B_{nm})$ and $(\gamma^C_{nm},\gamma^D_{nm})$ refer to the translational and chiral mode amplitudes of the active slip, respectively. 
$P^m_n(\cos \theta)$ denote associated Legendre polynomials of order $m$ and degree $n$. 
In the rest of the article, we denote $P^m_n(\cos\theta)$ as $P^m_n$.}

The boundary conditions at the surface of the swimmer and the far field read as
\begin{eqnarray}
\label{eq:bconditions}
\boldsymbol{u} \cdot \boldsymbol{n} &=& 0, \quad \text{on} \quad S, \label{eqn: impermeable}\\
\boldsymbol{u} \cdot \boldsymbol{t} &=& \left\{
\begin{array}{rl}
\displaystyle \boldsymbol{v_s} \cdot \boldsymbol{t} , & \text{on } \mho,\\
\displaystyle \frac{k}{\mu}(\boldsymbol{n}\cdot \boldsymbol{\sigma}) \cdot \boldsymbol{t}, & \text{on } S-\mho, 
\end{array} \right. \label{eqn: slip_condition}\\
\boldsymbol{u} &\to& -(\boldsymbol{U} + \boldsymbol{\Omega}\times\boldsymbol{r}) \quad \text{as} \quad \boldsymbol{r} \to \infty\, \label{eqn: farfield_condition}, 
\end{eqnarray}
where $\boldsymbol{n}, \boldsymbol{t}$ are, respectively, the unit normal and tangent vectors to the surface of the swimmer, $\boldsymbol{\sigma} = -p\boldsymbol{I} + \mu ({\boldsymbol{\nabla}} \boldsymbol{u}+{\boldsymbol{\nabla}} \boldsymbol{u}^T)$ is the stress tensor, where $\boldsymbol{I}$ is the identity matrix, and $T$ denotes transpose. $\boldsymbol{U}$ is the velocity and $\boldsymbol{\Omega}$ is the rotation rate of the swimmer. Here, $\boldsymbol{u}$ denotes the flow field in the body frame of reference. 
{Since different surface slips are prescribed over the patch $\mho$ and the rest of the surface $S-\mho$, the swimmer experiences discontinuous transitions of boundary conditions across these regions. 
This is analogous to the studies by Popescu et al.~\cite{popescu2018effective} and Yang et al.~\cite{yang2023stick}.} 


\section{Solution methodology}
\label{sec: Ana_solution}

The incompressible steady Stokes equation (Eqn.~\ref{eqn: stokes}), subject to the described boundary conditions, allows for analytical solutions to determine the swimmer's velocity and pressure fields. {While dealing with arbitrary Stokes flows, several authors used Lorentz reciprocal theorem to find the swimmer's velocity and rotation rate~\cite{stone1996propulsion,sellier2013arbitrary,yang2023stick} and also used Lamb's general solution~\cite{pak2014generalized}. Alternatively, for spherical geometries, the double curl method is widely used by several authors~\cite{padmavathi1993stokes,dhar2019self,biswas2023hydrodynamics}. In this approach, we represent the velocity $\boldsymbol{u}$ and pressure $p$ fields in terms of scalar functions as}
\begin{eqnarray}
\boldsymbol{u} &=& {\boldsymbol{\nabla}} \times {\boldsymbol{\nabla}} \times (\boldsymbol{r}A)+{\boldsymbol{\nabla}} \times (\boldsymbol{r}B), \label{eqn: dcu}\\
p &=& p_0+\mu\frac{\partial}{\partial r}(r{\boldsymbol{\nabla}}^2 A),\label{eqn: dcp}
\end{eqnarray}
where $\boldsymbol{r}$ is the position vector, $p_0$ is a constant, $A$, and $B$ are two independent scalar functions satisfying the biharmonic ${\boldsymbol{\nabla}}^4 A = 0$ and harmonic ${\boldsymbol{\nabla}}^2 B = 0$ equations.

\subsubsection{Solutions of $\mathbf{\nabla}^4 A=0$ and $\mathbf{\nabla}^2 B=0$}
\label{sec: flowfield}

As mentioned above, the scalar functions that satisfy the Stokes equation (Eqn.~\ref{eqn: stokes}) are the solutions of the biharmonic ${\boldsymbol{\nabla}}^4 A = 0$ and harmonic ${\boldsymbol{\nabla}}^2 B = 0$ equations. Results due to Almansi \cite{almansi1899sull}, we can express $A$ as $A = A_1 + r^2 A_2$, where $A_1$ and $A_2$ are two independent harmonics. As a result, the representation involves three independent scalar harmonic functions, namely $A_1, A_2$, and $B$. Therefore, the general solutions of ${\boldsymbol{\nabla}}^4 A = 0$ and ${\boldsymbol{\nabla}}^2 B = 0$ are given by
\begin{eqnarray}
A &=& \sum^\infty_{n=1} \Big[a_n r^n +b_n r^{n+2} + \frac{c_n}{r^{n+1}} + \frac{d_n}{r^{n-1}} \Big] S_n(\theta,\phi) \label{eqn: A}\\
B &=& \sum^\infty_{n=1} \Big[e_n r^n + \frac{f_n}{r^{n+1}} \Big] H_n(\theta,\phi), \label{eqn: B}
\end{eqnarray}
where $a_n,b_n,c_n,d_n,e_n,f_n$ are the unknown coefficients. 
$S_n(\theta,\phi) = \sum^n_{m=0} \big[A_{nm} \cos(m\phi) + B_{nm}\sin(m\phi)\big] P^m_n$, 
$H_n(\theta,\phi) = \sum^n_{m=0} \big[C_{nm}\cos(m\phi)+D_{nm}\sin(m\phi)\big] P^m_n$ are the spherical harmonics with $A_{nm},B_{nm},C_{nm},D_{nm}$ are the known coefficients. 

The same interfacial conditions (Eqs.~(\ref{eqn: impermeable}-\ref{eqn: slip_condition})) can be expressed in terms of $A$ and $B$, in the body frame of reference, as
\begin{eqnarray}
A &=& 0, \quad \text{on } S, \label{eqn: scalar_imp}\\
\frac{\partial A}{\partial r} &=&  \left\{
\begin{array}{rl}
&\displaystyle \sum_n \beta_n \tilde{S}_n(\theta,\phi), \quad \text{on } \mho, \\
&\displaystyle k \frac{\partial^{2} A}{\partial r^{2}},  \quad \text{on } S-\mho,
\end{array} \right.
 \\
B & = & \left\{
\begin{array}{rl}
&\displaystyle -\sum_n \gamma_n \tilde{H}_n(\theta,\phi), \quad \text{on } \mho, \\
&\displaystyle k \Big(\frac{\partial B}{\partial r}-\frac{B}{a}\Big), \quad \text{on } S-\mho,
\end{array} \right.
\end{eqnarray}
and the far-field condition (Eqn.~\ref{eqn: farfield_condition}) can be written as
\begin{eqnarray}
-u_1\boldsymbol{e_1}-u_2\boldsymbol{e_2}-u_3\boldsymbol{e_3} &=& 2a_1 (-A_{11}\boldsymbol{e_1}-B_{11}\boldsymbol{e_2}+A_{10}\boldsymbol{e_3}),\\
-\Omega_1\boldsymbol{e_1}-\Omega_2\boldsymbol{e_2}-\Omega_3\boldsymbol{e_3} &=& e_1 (-C_{11}\boldsymbol{e_1}-D_{11}\boldsymbol{e_2}+C_{10}\boldsymbol{e_3}).
\label{eqn: scalar_slip_patch_other}
\end{eqnarray}
where $(u_1,u_2,u_3),(\Omega_1,\Omega_2,\Omega_3)$ are the components of $\boldsymbol{U},\boldsymbol{\Omega}$ respectively {and $(\boldsymbol{e_1},\boldsymbol{e_2},\boldsymbol{e_3})$ denotes the unit vectors in the Cartesian coordinate system}. Therefore, the scalar fields in the body frame of reference by incorporating conditions Eqs.~(\ref{eqn: scalar_imp}-\ref{eqn: scalar_slip_patch_other}), due to the patch $\mho$ are given as
\begin{eqnarray}
    A_\mho &=& \Big[r  + \frac{a^3}{2 r^2} -\frac{3a}{2} \Big] \Big[-\frac{u_3}{2} P^0_1+\Big(\frac{u_1}{2} \cos \phi+\frac{u_2}{2}\sin \phi\Big)P^1_1\Big] 
    + \Big[\frac{a}{2}-\frac{a^3}{2r^2}\Big]\beta_1 \tilde{S}_1 \nonumber\\
    && \: + \sum^{\infty}_{n=2} \Big[ \frac{a^n}{2 r^{n-1}} - \frac{a^{n+2}}{2 r^{n+1}} \Big] \beta_n \tilde{S}_n, 
    \label{eqn: Aa}\\
    B_\mho &=& \Big[r - \frac{a^3}{r^2}\Big] \Big[(\Omega_1 \cos \phi+\Omega_2 \sin \phi)P^1_1-\Omega_3 P^0_1\Big]
    - \gamma_1 \Big(\frac{a}{r}\Big)^{2} \tilde{H}_1 - \sum^{\infty}_{n=2} \gamma_n\Big(\frac{a}{r}\Big)^{n+1}  \tilde{H}_n, \qquad \label{eqn: Ba} \quad 
\end{eqnarray}
whereas the same due to the rest of the surface $S-\mho$ is given as 
\begin{eqnarray}
   A_{S-\mho} &=& \Big[a - r +\frac{a^2 \,r^2 - a^4}{2 r^2 (a + 3 k)}\Big]
   \Big[\frac{u_3}{2} P^0_1 - \left(\frac{u_1}{2} \cos \phi+\frac{u_2}{2}\sin \phi\right) P^1_1\Big], 
   \label{eqn: Ap}\\
   B_{S-\mho} &=& \Big[r - \frac{a^4}{(a + 3 k)r^2}\Big] 
   \Big[-\Omega_3 P^0_1+\Big(\Omega_1 \cos \phi+\Omega_2 \sin \phi\Big)P^1_1\Big]. 
   \label{eqn: Bp}
\end{eqnarray}
The general solution of the velocity field $\boldsymbol{u}$ can be computed by implementing Eqs.~(\ref{eqn: Aa}-\ref{eqn: Bp}) to Eqn.~\ref{eqn: dcu}. Due to the restriction of the active region on the swimmer's surface, the expressions for the flow field, the velocity, and the rotation rate of the swimmer are cumbersome. Thus, we are not providing these expressions. However, we provide the velocity field $\boldsymbol{u}$ in the body frame of reference for the $n=1$ mode with the choice $\beta^A_{11} = 0 =\beta^B_{11}$ so that the reference fully covered swimmer translate along $z-$direction. Therefore, $\boldsymbol{u}$ due to the active patch $\mho$ given as
\begin{eqnarray}
    \boldsymbol{u}_{\mho} & = &\left(-1+\frac{3a}{2r} - \frac{a^3}{2 r^3} \right) \left( u_1\sin\theta \cos\phi + u_2\sin\theta \sin\phi +u_3\cos\theta \right)\boldsymbol{e_r} + \beta^A_{10}\bigg(\frac{a}{r} - \frac{a^3}{r^3} \bigg)\cos\theta \boldsymbol{e_r} \nonumber\\
    &&\: +\left(-1+ \frac{3a}{4r} + \frac{a^3}{4 r^3} \right) \Big[\big(u_1\cos\theta \cos\phi+u_2\cos\theta \sin\phi - u_3\sin\theta\big)\boldsymbol{e_\theta}- \big(u_1 \sin\phi - u_2\cos\phi\big)\boldsymbol{e_\phi}\Big] \nonumber\\
    &&\: - \beta^A_{10}\left(\frac{a}{2 r}+\frac{a^3}{2 r^3}\right) \sin \theta \boldsymbol{e_\theta} - \left(\frac{a^2}{r^2}\right) \Big[\big((a\Omega_1+\gamma^C_{11}) \sin \phi-(a\Omega_2+\gamma^D_{11}) \cos \phi \big)\boldsymbol{e_\theta} \nonumber\\
    &&\: +\big((a\Omega_1+\gamma^C_{11}) \cos\theta\cos\phi + (a\Omega_2+\gamma^D_{11}) \cos\theta\sin\phi- (a\Omega_3-\gamma^C_{10}) \sin\theta\big)\boldsymbol{e_\phi}\Big] \nonumber\\
    &&\: + r\Big[(\Omega_1\sin\phi - \Omega_2\cos\phi)\boldsymbol{e_\theta} - \big(\Omega_1 \cos\theta\cos\phi + \Omega_2 \cos\theta\sin\phi
- \Omega_3 \sin\theta\big)\boldsymbol{e_\phi}
\Big]
    \label{eqn: body_flow_field_active}    
\end{eqnarray}
and the same due to the rest of the surface $S - \mho$ given as,
\begin{eqnarray}
\boldsymbol{u}_{S-\mho} & = & \left(-1+\frac{3a(a+2k)}{2r(a+3k)}-\frac{a^4}{2r^3(a+3k)}\right)\big(u_1\sin\theta \cos\phi+u_2\sin\theta \sin\phi+u_3\cos\theta\big) \boldsymbol{e_r} \nonumber \\
&&\:+\left(-1+\frac{3a(a+2k)}{4r(a+3k)}+\frac{a^4}{4r^3(a+3k)}\right)
\Big[
\big(u_1\cos\theta \cos\phi+u_2\cos\theta \sin\phi-u_3\sin\theta\big)\boldsymbol{e_\theta} \nonumber \\
&&\: - \big(u_1 \sin\phi - u_2\cos\phi\big)\boldsymbol{e_\phi}\Big] + \left(r- \frac{a^4}{r^2(a+3k)}\right)\Big[(\Omega_1\sin\phi - \Omega_2\cos\phi)\boldsymbol{e_\theta} \nonumber \\
&&\: - \big(\Omega_1 \cos\theta\cos\phi + \Omega_2 \cos\theta\sin\phi
- \Omega_3 \sin\theta\big)\boldsymbol{e_\phi}
\Big].
\label{eqn: body_flow_field_passive}
\end{eqnarray}
{Note that in the rest of the article, we use ``fully covered swimmer" to denote a spherical swimmer with surface active slip velocity distributed uniformly over its entire surface.}


\subsection{Force-free and torque-free swimming}
\label{sec: F_T_free_swim}

The net drag $\boldsymbol{D}$ and net torque $\boldsymbol{T}$ of the swimmer can be obtained using the relations $\boldsymbol{D} = \int_\mho \boldsymbol{\sigma} \cdot \boldsymbol{n} ~ dS+\int_{S-\mho} \boldsymbol{\sigma} \cdot \boldsymbol{n} ~ dS$ and $\boldsymbol{T} =  \int_\mho \boldsymbol{r} \times (\boldsymbol{\sigma} \cdot \boldsymbol{n}) ~ dS+\int_{S-\mho} \boldsymbol{r} \times (\boldsymbol{\sigma} \cdot \boldsymbol{n}) ~ dS$, where $S$ denotes the surface of the swimmer. Since the swimmer is suspended in the quiescent fluid without any external forces and torques, its velocity $\boldsymbol{U}$ and rotation rate $\boldsymbol{\Omega}$ can be determined from the force-free ($\boldsymbol{D} = \boldsymbol{0}$) and torque-free ($\boldsymbol{T}=\boldsymbol{0}$) conditions. Correspondingly, $\boldsymbol{U}$ and $\boldsymbol{\Omega}$ as functions of angular coordinates, slip length, and modes of active slip velocity, can be represented mathematically as $\boldsymbol{U}=\boldsymbol{U}(\theta_1,\theta_2,\phi_1,\phi_2,k,\beta^A_{1m},\beta^B_{1m},\gamma^C_{1m},\gamma^D_{1m})$ and $\boldsymbol{\Omega}=\boldsymbol{\Omega}(\theta_1,\theta_2,\phi_1,\phi_2,k,\beta^A_{1m},\beta^B_{1m},\gamma^C_{1m},\gamma^D_{1m})$. {In the study of two faced slip-stick swimmer by Yang et al.~\cite{yang2023stick}, we see a similar structure for the velocity $\boldsymbol{U}$ as a function of slip partitioning angle, slip lengths, and the modes of active slip velocity. In our current study, in place of their slip partitioning angle, we have four angular parameters $(\theta_1,\theta_2,\phi_1,\phi_2)$ which determine the location of the patch}. Note that the corresponding expressions look cumbersome. Hence, we do not show their explicit forms here. As a limiting case, when $(\theta_1,\theta_2) = (0,\pi)$ and $(\phi_1,\phi_2) = (0,2\pi)$, i.e., the active slip is all over the surface of the fully covered swimmer, the corresponding velocity and rotation rate of the swimmer (with the choice $\beta^A_{11} = 0 =\beta^B_{11}$) in the laboratory frame of reference are given by
\begin{eqnarray}
    \boldsymbol{U_s} & = & -\frac{2\beta^A_{10}}{3}\boldsymbol{e_3}, \label{eqn: fully_vel}\\
    \boldsymbol{\Omega_s} & = & -\frac{\gamma^C_{11}}{a} \boldsymbol{e_1} -\frac{\gamma^D_{11}}{a} \boldsymbol{e_2} + \frac{\gamma^C_{10}}{a} \boldsymbol{e_3}, \label{eqn: fully_rot}
\end{eqnarray}
which agrees with the findings by Pak et al. \cite{pak2014generalized} and Burada et al. \cite{burada_pre}. Using the velocity, pressure, and stress fields, we can estimate the viscous power dissipation of the swimmer using the relation $P = -\int_\mho \boldsymbol{n} \cdot \boldsymbol{\sigma} \cdot \boldsymbol{u} ~ dS - \int_{S-\mho} \boldsymbol{n} \cdot \boldsymbol{\sigma} \cdot \boldsymbol{u} ~ dS$. Also, the Froude efficiency $\eta$ of the swimmer can be calculated as $\eta=(\boldsymbol{D_H}\cdot\boldsymbol{U}+\boldsymbol{T_H}\cdot\boldsymbol{\Omega})/P$, where $\boldsymbol{D_H}$ and $\boldsymbol{T_H}$ are the net hydrodynamic drag and torque, respectively, experienced by the swimmer given by Eqn.~\ref{eqn: NAxi_drag_H} and \ref{eqn: NAxi_torque_H}, in the absence of any active slip. Here, the corresponding expressions also look cumbersome, and hence, we do not show their explicit forms. Therefore, as a limiting case, the corresponding dissipated power and efficiency of the fully covered swimmer (with the choice $\beta^A_{11} = 0 =\beta^B_{11}$) are given by
\begin{eqnarray}
    P_s &=&  \frac{16}{3}a\pi\mu (\beta^A_{10})^2,\\
    \eta_s &=& \frac{8}{3} a \pi \mu \Big[(\beta^A_{10})^2 +  3 \Big((\gamma^C_{11})^2 + (\gamma^D_{11})^2 + (\gamma^C_{10})^2\Big)\Big]/P_s,
\end{eqnarray}
which agrees with the findings by Pak et al. \cite{pak2014generalized}.  {In this study, we set to $(\beta^A_{11},\beta^B_{11},\beta^A_{10})=(0,0,1)$ and $(\gamma^C_{11},\gamma^D_{11},\gamma^C_{10})=(1,1,1)/\sqrt{3}$ in such a way that the fully covered swimmer has magnitude of migration velocity $U_s=2/3$ and magnitude of rotation rate $\Omega_s=1$.}
 

\section{Symmetric active patch}
\label{sec: axi_model}

\begin{figure}
    \centering
    \includegraphics[width=0.8\linewidth]{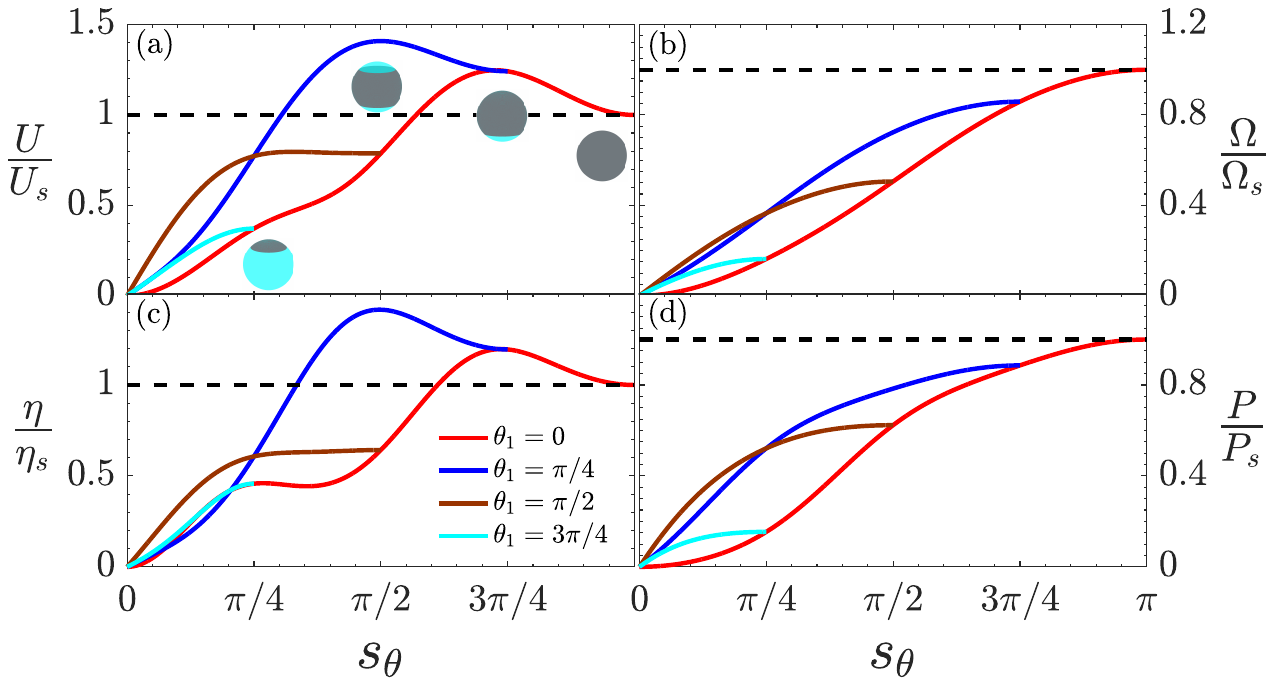}
    \caption{Variation of the strength of the velocity $U$ (a), the rotation rate $\Omega$ (b), the efficiency $\eta$ (c), and the dissipative power $P$ (d) of the swimmer with a symmetric patch (refer to figure~\ref{fig: schematic}(c)) as a function of $s_\theta = \theta_2 - \theta_1$ at different initial values of $\theta_1$. We choose $k/a = 0.01$, $(\beta^A_{11},\beta^B_{11},\beta^A_{10})=(0,0,1)$, and $(\gamma^C_{11},\gamma^D_{11},\gamma^C_{10})=(1,1,1)/\sqrt{3}$. $U_s, \Omega_s$ denote the strength of the velocity and rotation rate of the fully covered swimmer, respectively. {Whereas $\eta_s, P_s$ denote the efficiency and dissipated power of the fully covered swimmer, respectively}.}
    \label{fig: axi_vel_rot}
\end{figure}

We first consider the case of the swimmer with a symmetric active patch (refer to figure~\ref{fig: schematic}(c)) given by $\mho = \big\{ (r,\theta,\phi) \in S~|~ r = a,\, \theta_{1} \leqslant \theta \leqslant \theta_{2}, \, \phi_1 = 0, \,\phi_2 = 2 \pi \big\}$. 
Figure~\ref{fig: axi_vel_rot}(a) depicts the variation of the strength of the velocity $U$ with a polar patch width $s_\theta =\theta_2-\theta_1$ for different $\theta_1$ and fixed $k/a=0.01$. For $\theta_1=3\pi/4$, $U$ is monotonically increasing with $s-\theta$, but for $0 \leq \theta_1 < 3\pi/4$, it is non-monotonic and achieves maximum at $s_\theta=3\pi/4$ for $\theta_1=0$ whereas it is maximum at $s_\theta=\pi/2$ for $\theta_1=\pi/4$. Therefore, the strength $U$ is optimal for the patch configuration $(\theta^{o}_1,\theta^{o}_2)=(\pi/4,3\pi/4)$. Note that this optimal patch configuration is independent of $k$ and agrees with the finding obtained by Dhar et al. \cite{dhar2019self}. We now focus on the swimmer's efficiency (refer to figure~\ref{fig: axi_vel_rot}(c)). The optimal patch configuration that moves the swimmer with maximum velocity corresponds to maximum efficiency. Figure~\ref{fig: axi_vel_rot}(b) depicts the corresponding strength of the rotation rate $\Omega$ with $s_\theta$. With increasing $s_\theta$, the active region expands, and as a result, $\Omega$ increases for any configuration of the patch. However, $\Omega$ of a swimmer with the symmetric patch never exceeds $\Omega_s$, which corresponds to the strength of the rotation rate of a fully covered swimmer. Figure~\ref{fig: axi_vel_rot}(d) depicts the dissipative power $P$ with $s_\theta$. It is evident that $P$ increases as the active region expands for any patch configuration. This is due to increase in $\Omega$ (refer to figure~\ref{fig: axi_vel_rot}(b)). Note that the passive region, which is hydrophobic, counters the response of the active region. Hence, the swimmer with an active patch dissipates less power and never exceeds that of a fully covered swimmer (refer to figure~\ref{fig: axi_vel_rot}(d)). {It is noted that for other combinations of chiral modes $(\gamma^C_{11},\gamma^D_{11},\gamma^C_{10})$, the magnitude of rotation rate $\Omega$ vary, but the behavior of each $\Omega$ profile in Figure~\ref{fig: axi_vel_rot}(b) will remain the same and never exceeds that of a fully covered swimmer.}

\begin{figure}[b]
    \centering
    \includegraphics[width=\textwidth]{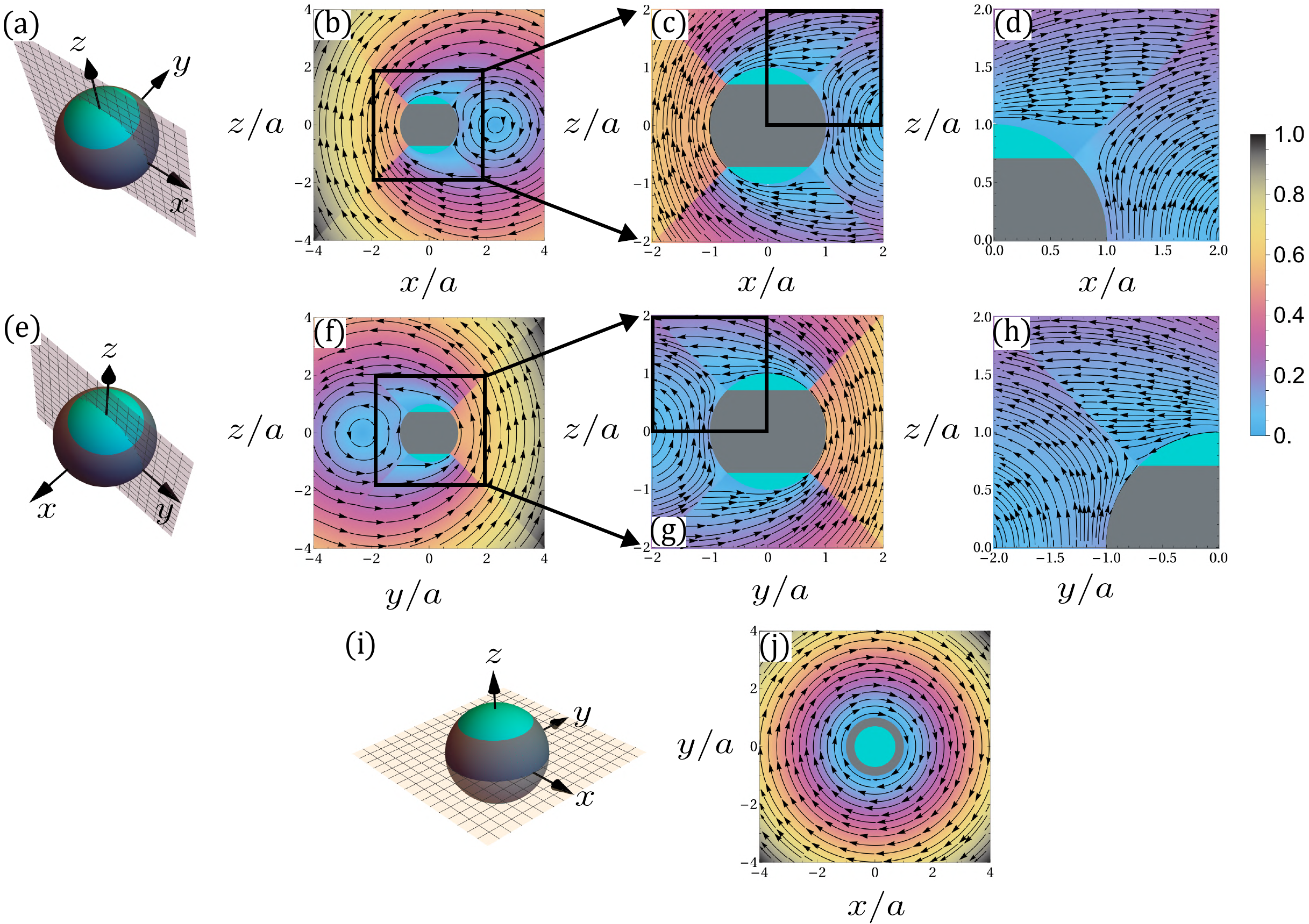}
    \caption{{The flow field generated by the swimmer with a symmetric patch for the optimal patch configuration $(\theta^{o}_1,\theta^{o}_2) = (\pi/4,3\pi/4)$, in the body frame of reference. The flow field profiles are projected onto the $xz$ plane (a-d), on the $yz$ plane (e-h), and on the $xy$ plane (i-j). The planes intersect the swimmer (indicated in cyan), with the symmetric patch (represented by the grey region) displayed in sub-figures (a, e, i). The discontinuity in the magnitude of the perturbed flow, caused by the discontinuity at the interface between the patch $\mho$ and the rest of the surface $S-\mho$, is shown in sub-figures (b, f, j). Correspondingly, the deflection in the streamlines near the swimmer is shown in enlarged sub-figures (c-d, g-h) for the respective planes. The other parameters are the same as in figure~\ref{fig: axi_vel_rot}.}}
    \label{fig: axi_FF}
\end{figure}
{
The expression for the velocity field $\boldsymbol{u}$ in the body frame of reference looks cumbersome, so we are not providing it here. However, we present the graphical results. The general velocity field $\boldsymbol{u}$ can be expressed as $\boldsymbol{u} \sim \boldsymbol{\mathcal{A}}r^{-1}+\boldsymbol{\mathcal{B}}r^{-2}+\boldsymbol{\mathcal{C}}r^{-3}$, where, $\boldsymbol{\mathcal{A}}=\boldsymbol{\mathcal{A}}(\theta,\phi,k,\boldsymbol{U}), \boldsymbol{\mathcal{B}}=\boldsymbol{\mathcal{B}}(\theta,\phi,k,\boldsymbol{\Omega})$ and $\boldsymbol{\mathcal{C}}=\boldsymbol{\mathcal{C}}(\theta,\phi,k,\boldsymbol{U})$ are the amplitudes which are comparable for $(\theta^{o}_1,\theta^{o}_2) = (\pi/4,3\pi/4)$. It is evident that $\boldsymbol{u}$ encompasses Stokeslet and source dipole components, corresponding to $\sim r^{-1}$ and $\sim r^{-3}$ terms, respectively, and a Rotlet component corresponding to $\sim r^{-2}$. 
When a fully covered swimmer moves with a velocity of $\boldsymbol{U_s}$ and a rotation rate of $\boldsymbol{\Omega_s}$, the resulting flow satisfies force-free and torque-free conditions, characteristic of free-swimming motion \cite{pak2014generalized}. As a result, the Stokeslet and Rotlet components do not contribute to the velocity field $\boldsymbol{u}$. 
However, in the current study, the surface activity concentrates only on the patch $\mho$, while the rest of the surface $\mathcal{S}-\mho$ is hydrophobic with a slip length $k\ll a$. As the swimmer moves with velocity $\mathbf{U}$ and rotation rate $\mathbf{\Omega}$, the Stokeslet (also Rotlet) components do not cancel each other due to the dependency of $\mathbf{U}$ and $\mathbf{\Omega}$ on the patch configurations and the slip length $k/a$. 
Consequently, Stokeslet and Rotlet components appear in the flow field $\boldsymbol{u}$.
One can see the presence of these singularities in the vicinity of the swimmer and vanish at the far field. We Justify this in detail in Appendix~\ref{sec: hyd_sing}.}

{
To capture the impact of the boundary data discontinuity in the flow field near the surface, we analyze the flow field patterns in the $xz,yz$ and $xy$ planes in the body frame of reference of the swimmer, with the optimal symmetric patch configuration $(\theta_1^o, \theta_2^o) = (\pi/4, 3\pi/4)$ which is depicted by figure~\ref{fig: axi_FF}. Figure~\ref{fig: axi_FF}(a, e) illustrates the location of $xz$ and $yz$ planes that cut the swimmer with a symmetric patch vertically. 
Note that, unlike the fully covered swimmer, the flow field over $xz$ and $yz$ planes are not mirror symmetric due to the presence of different surface properties and chirality. Figure~\ref{fig: axi_FF}(b, f) depicts that, due to the inhomogeneous boundary conditions over the surface of the swimmer, the magnitude of the flow field displays a noticeable jump due to the presence of the patch $(\theta_1^o, \theta_2^o) = (\pi/4, 3\pi/4)$. Correspondingly, one can see the jump along the lines $\theta=\pi/4$ and $\theta=3\pi/4$. However, this jump is prominent only in the vicinity of the swimmer and diminishes at the far field. 
The presence of a patch deflects the streamlines which are otherwise continuous in the entire domain. These deflections are more visible in the vicinity of the swimmer along the lines $\theta=\pi/4$ and $\theta=3\pi/4$ and show no deviations at far field. The $xy$ plane cuts the swimmer horizontally and contains the effect of the patch only (refer to figure~\ref{fig: axi_FF}(i)). Therefore, no discontinuity in the magnitude of the flow field and deviations in streamlines were observed in the vicinity of the swimmer as well as at the far field. However, in this case, the flow field pattern is mirror symmetric due to the symmetry of the swimmer with respect to the $z$ axis (refer to Figure \ref{fig: axi_FF}(j)).}

\section{Arbitrary active patch}
\label{sec: Naxi_model}
\begin{figure}[b]
    \centering
    \includegraphics[width=1\textwidth]{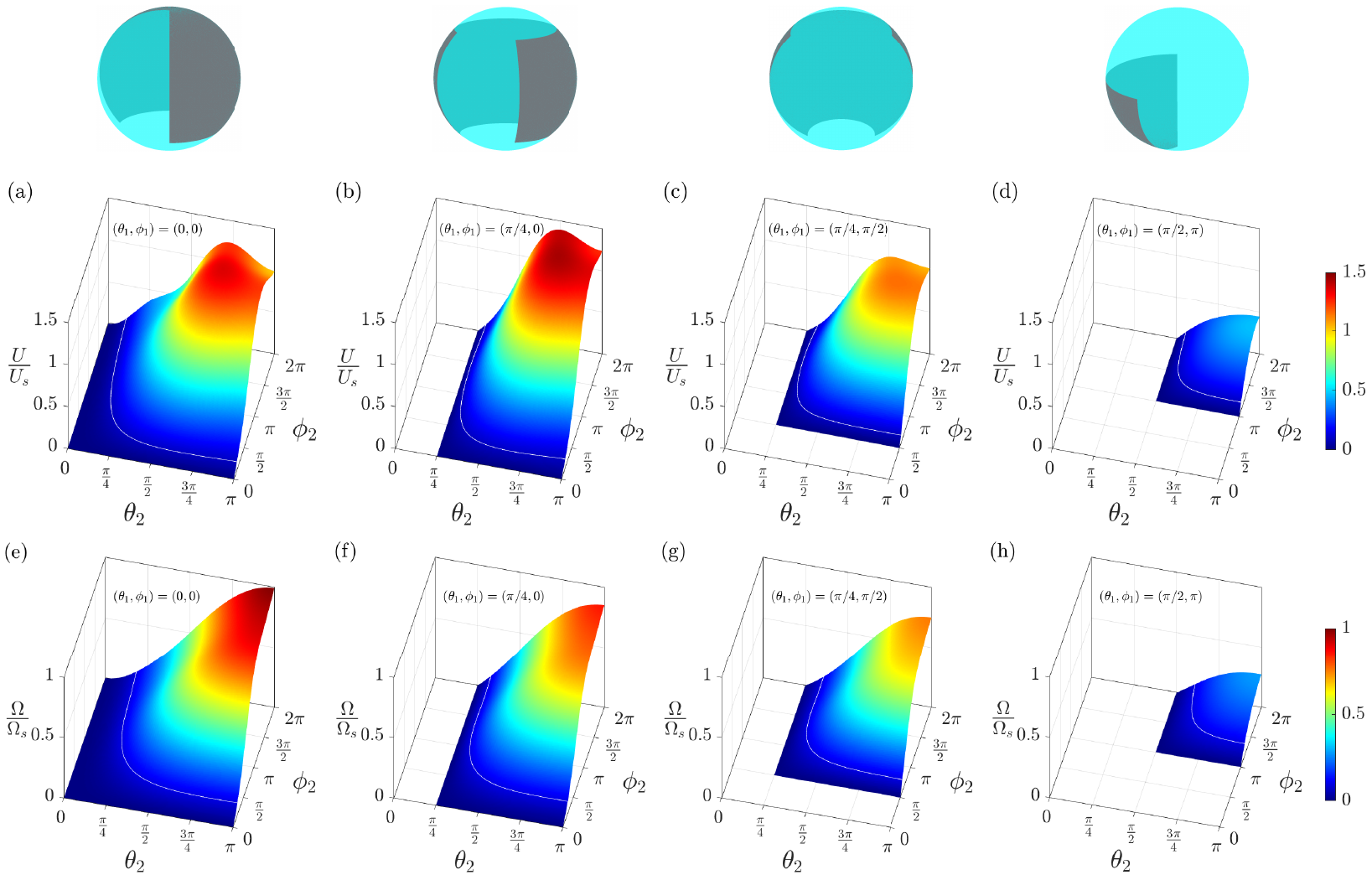} 
    \caption{Variation of the strength of the velocity $\boldsymbol{U}$ and the rotation rate $\boldsymbol{\Omega}$ of the swimmer as a function of $\theta_2$ and $\phi_2$ for a particular choice of $(\theta_1,\phi_1)$. Corresponding local optimal configurations of the patch have been shown on the top panel. The other parameters are the same as in figure~\ref{fig: axi_vel_rot}.}
    \label{fig: opt_all}
\end{figure}

Here, we consider a swimmer with a hydrophobic surface on the region $S-\mho$, and the activity is on an arbitrary patch region $\mho = \big\{ (r,\theta,\phi) \in S~|~ r=a,\theta_{1} \leqslant \theta \leqslant \theta_{2}, \phi_1 \leqslant \phi \leqslant \phi_2 \big\}$. For a given slip length $k/a$ on the hydrophobic region $S-\mho$, we control the active patch $\mho$ by fixing the polar and azimuthal angles $(\theta_1,\phi_1)$ while varying $(\theta_2,\phi_2)$. Figure \ref{fig: opt_all} depicts the variation in $U$ and $\Omega$ with $(\theta_2,\phi_2)$. For a fixed $\theta_2$, the swimmer experiences no significant velocity and rotation until the threshold pair of $(\theta_2,\phi_2)$. The locus of such threshold pair $(\theta_2,\phi_2)$ lies on the equimagnitude curves $U=0.1$ and $\Omega=0.1$ shown as a white contour in figure~\ref{fig: opt_all}. Further increasing the size of the patch, the swimmer attains maximum velocity at a specific $(\theta_2,\phi_2)$. Nevertheless, it is important to note that this particular trend does not persist, as further increasing $(\theta_2,\phi_2)$, there is a reduction in the velocity. Similarly, the strength of the rotation rate also increases beyond the threshold patch. However, the rotation rate due to an arbitrary active patch never exceeds $\Omega_s$ that of a fully covered swimmer (refer to figure~\ref{fig: opt_all}(e-h)), similar to the symmetric patch case. Accordingly, we have the optimal configuration leading to the maximum value of $U$ for a fixed $(\theta_1,\phi_1)$ as the local optimal configuration. One can realize that for a fixed $k/a$, the optimal value of $U$ and $\Omega$ not only depend on $(\theta_2,\phi_2)$ but also depend on $(\theta_1,\phi_1)$. This means the location of the active patch on the surface of the swimmer controls this optimal configuration.

\subsection{Arbitrary patch with a fixed surface area}

\begin{figure}[b]
    \centering
    \includegraphics[width=1\textwidth]{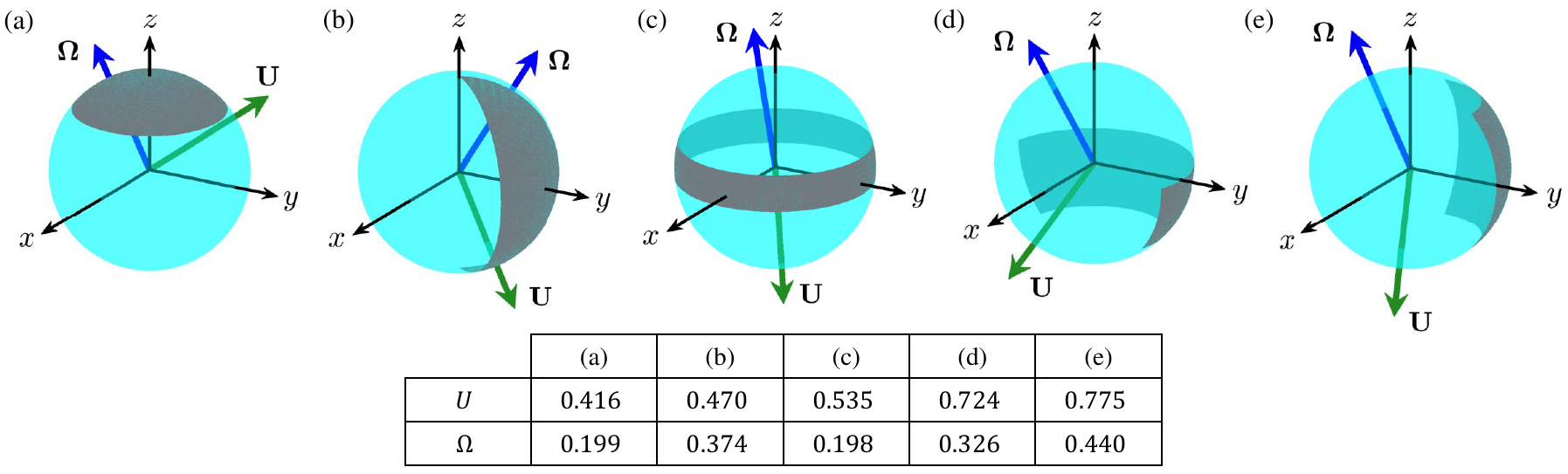} 
    \caption{Variation of the velocity $\boldsymbol{U}$ and the rotation rate $\boldsymbol{\Omega}$ of the swimmer with various configurations of the patch of fixed area. The strength of $\boldsymbol{U}$ and $\boldsymbol{\Omega}$ are provided in the tabular form, and their {position vectors} are depicted by green and blue arrows, respectively. The other parameters are the same as in figure~\ref{fig: axi_vel_rot}.}
    \label{fig: rotation}
\end{figure}
To understand the dependency of $(\theta_1,\theta_2,\phi_1,\phi_2)$ on the optimal configuration, we consider a test case $(\theta_1,\theta_2,\phi_1,\phi_2)=(0,0.88,0,2\pi)$ that corresponds to a patch ({cap at the north pole}) with a fixed surface area (refer to figure~\ref{fig: rotation}(a)). We now vary the shape and location of the patch, keeping the area fixed. For brevity, we show four such configurations (refer to figure~\ref{fig: rotation}(b-e)). For each of these configurations (the area is fixed), the directions of velocity $\boldsymbol{U}$ and rotation rate $\boldsymbol{\Omega}$ vary as depicted in figure~\ref{fig: rotation}(a-e). We have computed corresponding strengths $U$ and $\Omega$ as shown in the table in figure~\ref{fig: rotation}. Interestingly, the location and shape of the patch play a vital role in controlling the swimmer's locomotion. Looking at figure~\ref{fig: rotation}(a) ({cap at the north pole}) and figure~\ref{fig: rotation}(c) (strip at the equator), we infer that the latter generates higher velocity than the former. However, the swimmer can achieve maximum velocity due to an arbitrary active patch at different locations (refer to figure~\ref{fig: rotation}(d-e)) among the chosen configurations. These observations may help experimentalists design swimmers with controlled locomotion by varying the shape and location of the active patch. After analyzing the local optimal configuration that gives maximum $\boldsymbol{U}$ for fixed $(\theta_1,\phi_1)$, {it is interesting to ask which general configuration (arbitrary active patch) leads to the maximum velocity or efficiency of the swimmer by varying $(\theta_1,\theta_2,\phi_1,\phi_2)$ for fixed $k/a$. To answer this question, we solve an optimization problem.}


\subsection{Global optimal configuration}
\label{subsec: opt_conf}

\begin{figure*}[h]
    \centering
    \includegraphics[width=\linewidth]{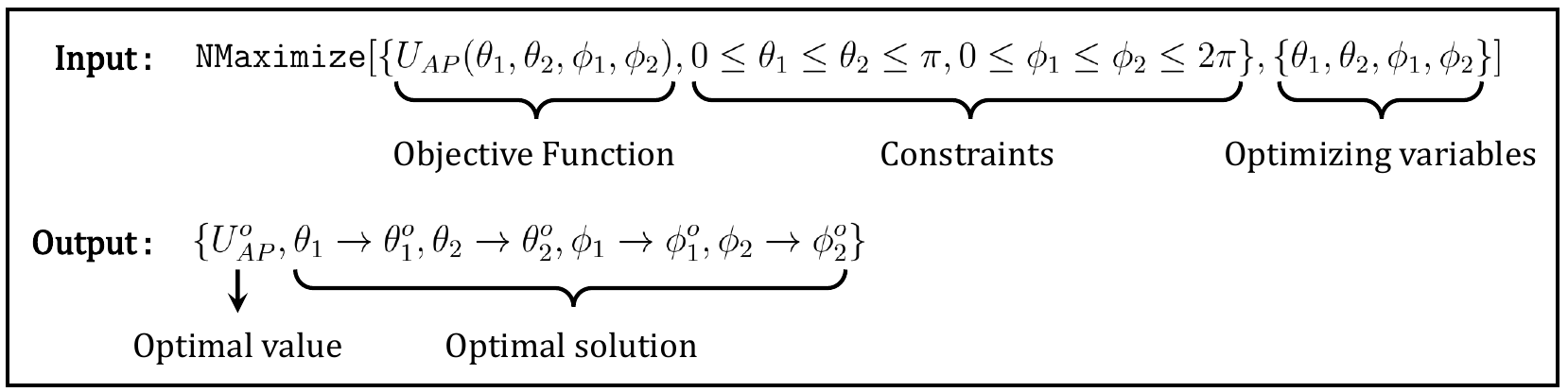}
    \caption{Pseudo-code of Mathematica 13.2 in-built function \texttt{NMaximize}.}
    \label{fig: pseudo}
\end{figure*}

To find the global optimal configuration, we choose the active slip parameters $(\beta^A_{11},\beta^B_{11},\beta^A_{10})=(0,0,1)$ and $(\gamma^C_{11},\gamma^D_{11},\gamma^C_{10})=(1,1,1)/\sqrt{3}$ as before. Let $U_{AP}=U_{AP}(\theta_1,\theta_2,\phi_1,\phi_2)$ denotes the strength of the velocity of a swimmer with an arbitrary patch for fixed $k\ll a$. With this $U_{AP}$ as the objective function, we define the following optimization problem
\begin{empheq}[box=\fbox]{align}
    \text{Maximize} &\quad U_{AP}(\theta_1,\theta_2,\phi_1,\phi_2), \label{eqn: opt_prob_e} \\
    \text{subject to,} &\quad 0 \leq \theta_1 \leq \theta_2 \leq \pi,    \quad 0 \leq \phi_1 \leq \phi_2 \leq 2\pi. \label{eqn: opt_prob}
\end{empheq}
{The expression for $U_{AP}$ is a nonlinear function of $(k, \theta_1, \theta_2, \phi_1, \phi_2)$, which is cumbersome and difficult to handle analytically. To address this, we employed numerical nonlinear global optimization tools available in Mathematica 13.2. The built-in function {\texttt{NMaximize}} was used with prescribed constraints over the entire domain to find the global maximum. The {\texttt{NMaximize}} function attempts to find the global maximum of the objective function subject to the given constraints. Based on the nature of the objective function, {\texttt{NMaximize}} automatically selects an in-built method such as Differential Evolution, Nelder-Mead, Random Search, or Simulated Annealing. Since the objective function $U_{AP}$ in this study is sufficiently differentiable, the ``Differential Evolution" method can be specifically applied to find the optimal solution $(\theta_1^o, \theta_2^o, \phi_1^o, \phi_2^o)$ and the corresponding optimal value of $U_{AP}$. For a fixed $k \ll a$, the pseudo-code for this function is outlined as in figure~\ref{fig: pseudo}.}

\begin{table}[h]
\caption{\label{tab:opt_mig_table}%
Variation of the strength of the optimal velocities and the corresponding optimal configurations of the patch with $k/a$. $U^{o}_{SP}$ and $U^{o}_{AP}$ correspond to the strength of the optimal velocity of the swimmer with the symmetric and arbitrary patches, respectively, scaled with $U_s$. $\Omega_{SP}$ and $\Omega_{AP}$ correspond to the strength of the rotation rate of the swimmer with the symmetric and arbitrary patches, respectively, scaled with $\Omega_s$.}

\begin{ruledtabular}
    \begin{tabular}{cc|ccc|ccccc}
    ~ & ~ & \multicolumn{2}{c}{Symmetric Patch} & ~ & \multicolumn{5}{c}{Arbitrary Patch}\\[0.5ex]
    ~ & ~ &  \multicolumn{2}{c}{$\displaystyle \theta^{o}_1=\frac{\pi}{4},~\theta^{o}_2=\frac{3\pi}{4}$} & ~ & \multicolumn{5}{c}{$ \displaystyle \theta^{o}_1=\alpha\,\frac{\pi}{4},~\theta^{o}_2=\delta\,\frac{3\pi}{4},~\phi^{o}_1=0,~\phi^{o}_2=\lambda\,\frac{3\pi}{2}$}\\[0.5ex]
    $k/a$ & ~ & \multicolumn{1}{c}{$U^{o}_{SP}$} & \multicolumn{1}{c}{$\Omega_{SP}$} & ~ & \multicolumn{1}{c}{$\alpha$} & \multicolumn{1}{c}{$\delta$} & \multicolumn{1}{c}{$\lambda$} & \multicolumn{1}{c}{$U^{o}_{AP}$} & \multicolumn{1}{c}{$\Omega_{AP}$}\\ 
    \hline
    $0$	   & ~ &	$1.414$   & $0.718$ & ~ &   $0.894$   &   $1.037$   &  $1.012$  &   $1.450$ & $0.714$ \\[0.5ex]
    $0.02$ & ~ &	$1.403$   & $0.728$ & ~ &   $0.906$   &   $1.034$   &  $1.018$  &   $1.433$ & $0.723$\\[0.5ex]
    $0.04$ & ~ &	$1.394$   & $0.738$ & ~ &   $0.917$   &   $1.032$   &  $1.026$  &   $1.416$ & $0.730$\\[0.5ex]
    $0.06$ & ~ &	$1.385$   & $0.747$ & ~ &  $0.927$   &   $1.029$   &  $1.033$  &   $1.400$ & $0.735$\\[0.5ex]
    $0.08$ & ~ &	$1.377$   & $0.755$ & ~ &  $0.936$   &   $1.027$   &  $1.039$  &   $1.386$ & $0.739$\\[0.5ex]
    $0.1$  & ~ &	$1.371$   & $0.763$ & ~ &  $0.944$   &   $1.025$   &  $1.046$  &   $1.373$ & $0.743$\\
  \end{tabular}
\end{ruledtabular}
\end{table}

For fixed $k \ll a$, we see that the solution of the above optimization problem Eqs.~(\ref{eqn: opt_prob_e}-\ref{eqn: opt_prob}) is unique. Corresponding to this optimal value of $U_{AP}$ which we denoted by $U^{o}_{AP}$, we identify the patch configuration by $(\theta^{o}_1,\theta^{o}_2,\phi^{o}_1,\phi^{o}_2)$. In Table \ref{tab:opt_mig_table}, we present the optimal velocities of the swimmer for symmetric patch $(U^{o}_{SP})$ and arbitrary patch $(U^{o}_{AP})$, for different $k/a \ll 1$ in the range $(0,0.1)$. Unlike the symmetric patch model, the optimal configuration for an arbitrary patch depends on $k/a$. Accordingly, we have introduced parameters $\alpha,\delta, \lambda$ to show these variations. It is observed that swimmers with higher hydrophobicity (larger $k/a$) experience smaller velocities both for symmetric and arbitrary active patches. However, for the entire range of $k/a \ll 1$ considered, $U^{o}_{AP}$ is always higher than $U^{o}_{SP}$. However, when $k/a \sim 0.1$, we observe that $U^{o}_{AP} \sim U^{o}_{SP}$. It means that the type of active patch, i.e., symmetric or arbitrary, does not influence the velocity much when the rest of the surface ($S-\mho$) is highly hydrophobic. On the other hand, when the rest of the surface is hydrophilic ($k/a = 0$), then the difference between $U^{o}_{AP}$ and $U^{o}_{SP}$ is maximum. This may have a significant importance in designing artificial swimmers with materials having varied surface roughness. Furthermore, we calculate the efficiency of the swimmer for optimal configurations in both symmetric and arbitrary patch models. Our observations reveal that when the rest of the surface is hydrophilic $(k/a = 0)$ or hydrophobic up to a certain threshold value of $(k/a)$, the swimmer's efficiency with an arbitrary patch $(\eta_{AP})$ surpasses that of a swimmer with a symmetric patch $(\eta_{SP})$. However, beyond this threshold value of $(k/a)$, due to $U^{o}_{AP} \sim U^{o}_{SP}$, the swimmer with a symmetric patch demonstrates greater efficiency than the swimmer with an arbitrary patch.

\begin{figure}
    \centering
    \includegraphics[width=\textwidth]{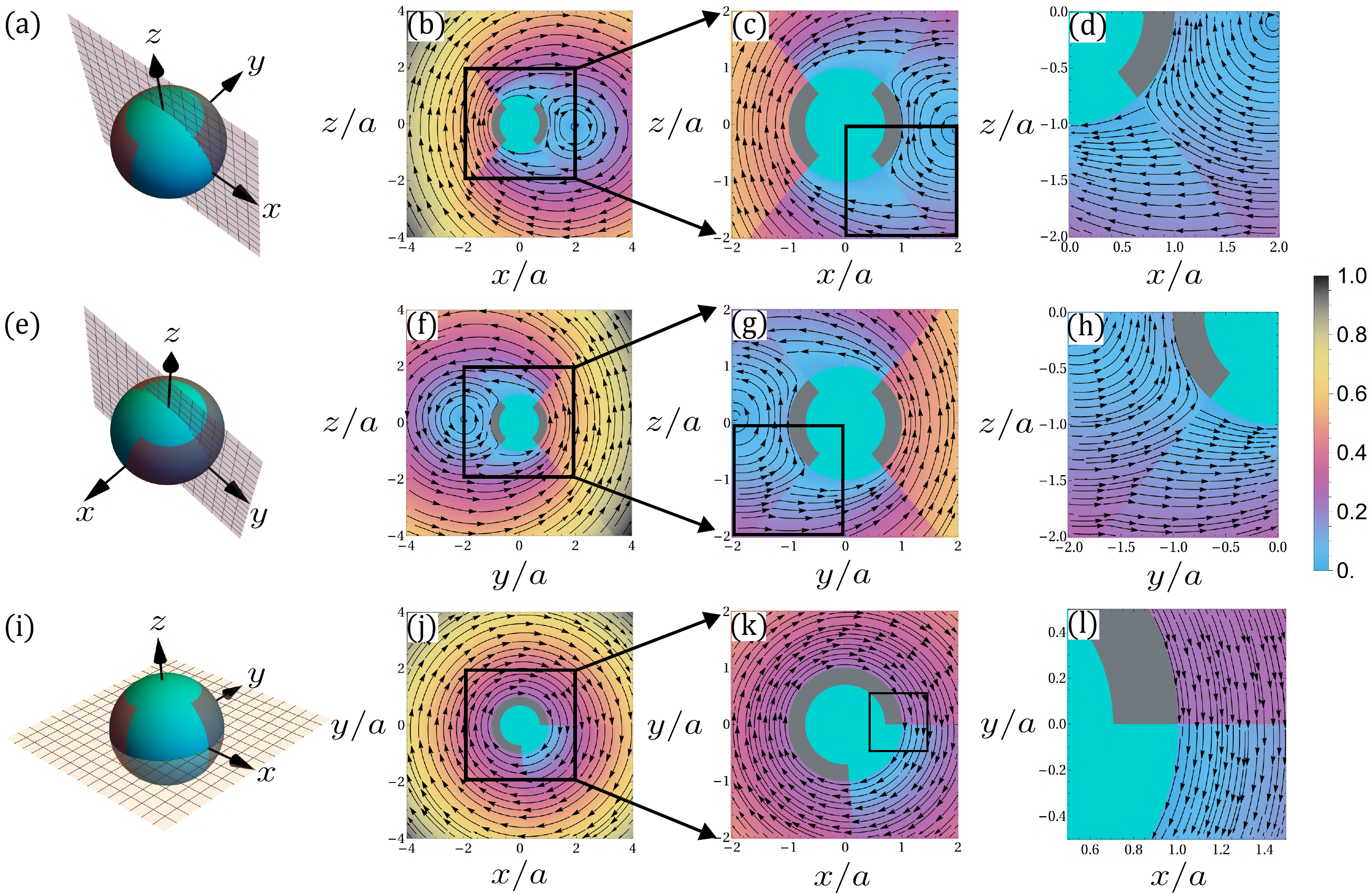}
    \caption{{The flow field generated by the swimmer with the global optimal configuration $(\theta^{o}_1,\theta^{o}_2,\phi^{o}_1,\phi^{o}_2)=(0.899\times\pi/4, 1.036\times 3\pi/4, 0, 1.015\times 3\pi/2)$ corresponding to $k/a = 0.01$, in the body frame of reference.  The flow field profiles are projected onto the $xz$ plane for $y=0.5$ (a-d), on the $yz$ plane (e-h), and on the $xy$ plane (i-j) to capture the effect of patch $\mho$ and rest of the surface $S-\mho$. The planes intersect the swimmer (indicated in cyan), with the symmetric patch (represented by the grey region) displayed in sub-figures (a, e, i). The propagation of the perturbed flow, caused by the discontinuity at the interface between the patch $\mho$ and the rest of the surface $S-\mho$, is shown in enlarged sub-figures (c-d, g-h, k-l) for the respective planes. The other parameters are the same as in figure~\ref{fig: axi_vel_rot}.}}
    \label{fig: Naxi_FF}
\end{figure}

{
In Figure~\ref{fig: Naxi_FF}, we present the flow fields generated by the swimmer with an arbitrary patch however with an optimal configuration $(\theta^{o}_1,\theta^{o}_2,\phi^{o}_1,\phi^{o}_2)=(0.899\times\pi/4, 1.036\times 3\pi/4, 0, 1.015\times 3\pi/2)$, corresponding to $k/a = 0.01$, in the body frame of reference across different planes. To capture the effect of the patch and the rest of the surface on the flow field patterns, we examine the $xz,yz$, and $xy$ planes in the body frame of the swimmer, as shown in Figure~\ref{fig: Naxi_FF}(a, e, i). These planes reflect the influence of the patch $\mho$ and the rest of the surface $S-\mho$, corresponding to regions governed by $\boldsymbol{u}_\mho$ and $\boldsymbol{u}_{S-\mho}$, respectively. Figure~\ref{fig: Naxi_FF}(b, f) shows that the optimal $(\theta^{o}_1,\theta^{o}_2)=(0.899\times\pi/4, 1.036\times 3\pi/4)$ is close to $(\pi/4,3\pi/4)$, hence the flow field patterns in the $xz$ and $yz$ planes resemble those of a swimmer with a symmetric patch (figure~\ref{fig: axi_FF}(b, f)). However, due to the incomplete coverage of patch along azimuthal angle $\phi$, i.e., $(\phi^{o}_1,\phi^{o}_2)=(0,1.015\times 3\pi/2)$, the centers of the vortices in both planes are shifted away from the corresponding axes as opposed to the case of symmetric patch (figure~\ref{fig: axi_FF}). However, the behavior of the flow fields remains the same as in the symmetric patch case, with the jump in the magnitude of the flow field along the lines $\theta=0.899\times\pi/4$ and $\theta=1.036\times3\pi/4$.
Unlike the symmetric patch case, the $xy$ plane in this configuration shows the influence of both the patch and the rest of the surface (refer to figure~\ref{fig: Naxi_FF}(i)). As a result, a discontinuity in the magnitude of the flow field is visible near the swimmer, but it rapidly vanishes in the far field. The streamlines remain continuous with no significant deviations observed near the swimmer along the lines $\phi=0$ and $\phi=1.015\times3\pi/2$, which are influenced by the patch (refer to figure~\ref{fig: Naxi_FF}(k-l)).
}


\section{Discussion}
\label{sec: discussion}

This work focuses on the mathematical modeling of the locomotion of self-propelled particles to explore their controllability and efficiency. Now, we wish to refer to a few experimental works on swimming microorganisms and bring insights into the context of the present study. {These studies involve various parameters like viscosity, the radius of the swimmer, and surface active slip parameters, which are chosen based on the experimental situations}. Our theoretical model also involves each parameter and hence may serve as an alternative tool to generate different physical conditions. In this context, we refer to the work by McConnell et al. \cite{mcconnell2010patchy} and Xu et al. \cite{xu2011increasing}, where they developed a novel method for fabricating patchy, gold-on-silica Janus particles and reported advanced synthetic techniques for preparing silica nanoparticles with rough surfaces in a reverse micro-emulsion followed by drying treatment, respectively. 
We try to mimic such models to establish that the results developed in this current work serve as tools to understand a larger class of physical models for which experimental results are unavailable. We consider rough-surfaced silica Janus particles with a partial gold patch coating. The typical radius of such a particle is $a \sim 300$ nm \cite{lu2003asymmetric}. We consider a hypothetical surface active slip velocity given by $(\beta^A_{11},\beta^B_{11},\beta^A_{10})=(0,0,1.5)$ nm/s and $(\gamma^C_{11},\gamma^D_{11},\gamma^C_{10})=a(1,1,1)/\sqrt{3}$ nm/s prescribed over the arbitrary gold patch configuration defined by $(\theta_1,\theta_2,\phi_1,\phi_2)=(\pi/4, 3\pi/4, 0, 3\pi/2)$ and rest of the surface is with slip length $k \sim 10$ nm. Suppose that, the particle is suspended in 30 wt\% $\text{H}_2\text{O}_2$ solution of viscosity $\eta \sim 1.08$ mPa$\cdot$s at temperature $\text{T} \sim 293$K. We realize for these prescribed surface properties, the strength of the velocity of the particle with the prescribed gold patch is $\sim 84$ nm/s. Predictions based on these realistic values are likely to aid experimentalists in developing effective spherical active swimmers with an arbitrary active patch for the targeted and controlled release of drugs.


\section{Conclusions}
\label{sec: conclusions}

In this work, we have considered a spherical swimmer of radius $a$ with an arbitrary active patch with the surface slip velocity $\boldsymbol{v_s}$ suspended in a quiescent ambient flow. The rest of the surface of the swimmer is inactive and is hydrophobic with a {Navier slip} of length $k \ll a$. The velocity $\boldsymbol{U}$, rotation rate $\boldsymbol{\Omega}$, and the flow field $\boldsymbol{u}$ (in the body frame of reference) are calculated analytically for the swimmer with a symmetric and arbitrary patch. The main objective of this work is to find an efficient optimal configuration of the patch for which the strength of the velocity is maximum. This study also illustrates how $k$, $\boldsymbol{v_s}$, and angular widths play a key role in controlling the locomotion of the swimmer. In the symmetric patch model, we find the optimal symmetric patch configuration $(\theta^{o}_1,\theta^{o}_2)=(\pi/4,3\pi/4)$, which is independent of $k$, gives maximum strength of the velocity. It is observed that for $k = 0.01$, we have $U^{o}_{SP}=1.409~U_s$, where $U_s$ denotes the strength of the velocity of a fully covered swimmer (activity all over the surface). Moreover, the swimmer is more efficient for this optimal patch configuration than any other symmetric patch configuration. In particular, $\eta^{o}_{SP} \approx \sqrt{2}~\eta_s$, where $\eta_s$ denotes the efficiency of the fully covered swimmer. Unlike the symmetric patch model, for the arbitrary patch model, the optimal patch configuration to achieve optimal strength of the velocity $U^{o}_{AP}$ shows a dependency on $k$. We declare that this optimal arbitrary patch configuration is globally optimal compared to symmetric and asymmetric patch models. In particular, for $k/a=0.01$, we have $U^{o}_{SP}(\approx 1.409~U_s)<U^{o}_{AP}(\approx 1.441~U_s)$. In both patch configurations, the dissipated power of the swimmer is bounded above by the dissipated power of the fully covered swimmer. {Due to the inhomogeneous boundary conditions over the surface of the swimmer, a discontinuity in the magnitude of the flow field is visible near the swimmer, but it vanishes in the far field. The streamlines are also impacted by the presence of the patch. The patch deflects the streamlines which are otherwise continuous in the entire domain. These deflections are visible near the swimmer and show no deviations at the far field}. We anticipate our research will inspire experimentalists to develop effective, well-controllable artificial microswimmers.


\newpage
\appendix

\section{Presence of hydrodynamic singularities in the vicinity of the swimmer}
\label{sec: hyd_sing}

{In Section~\ref{sec: flowfield}, we provided the general solution to the Stokes equations (Eqn.~\ref{eqn: stokes}), subject to the prescribed boundary conditions Eqs.~(\ref{eqn: impermeable}-\ref{eqn: slip_condition}) in the body frame of reference, as expressed by $\boldsymbol{u}_\mho$ (refer to Eqn.~\ref{eqn: body_flow_field_active}) due to the patch region $\mho$. Without loss of generality, we consider a swimmer with a symmetric patch, and over this, we set active slip's translational mode $\beta^A_{10} = \beta$ and assume that it has chirality with mode $\gamma^C_{10} = \gamma$ and all other modes are set to zero. Consequently, applying force-free and torque-free conditions, the swimmer's velocity $\boldsymbol{U}$ and rotation rate $\boldsymbol{\Omega}$ are given as
\begin{equation}
    \boldsymbol{U} = \underbrace{-(a+3k) \frac{2\beta(3\Theta_1-\Theta_3)}{12(a+2k)+3k(3\Theta_1-\Theta_3)}}_{\textrm{$u_3$}}\boldsymbol{e_3}, \quad \boldsymbol{\Omega} = \underbrace{\bigg(\frac{a+3k}{a}\bigg) \frac{\gamma(9\Theta_1-\Theta_3)}{16a+3k(9\Theta_1-\Theta_3)}}_\textrm{$\Omega_3$}\boldsymbol{e_3},
    \label{eqn: u3_o3}
\end{equation}
where $\Theta_i = \cos i\theta_2-\cos i\theta_1$ for $i=1,2,3$. As a result, expressions for $\boldsymbol{u}_\mho$ simplifies to
\begin{eqnarray}
    \boldsymbol{u}_{\mho} & =& -\boldsymbol{U}-\boldsymbol{\Omega}\times \boldsymbol{r} +\frac{a}{4r}(3u_3+2\beta)(2\cos\theta \boldsymbol{e_r}-\sin\theta\boldsymbol{e_\theta}) + \left(\frac{a^2}{r^2}\right) (a\Omega_3-\gamma) \sin\theta\boldsymbol{e_\phi}\nonumber\\
    &&\:- \frac{a^3}{4r^3}(u_3+2\beta)(2\cos \theta \boldsymbol{e_r}+\sin\theta\boldsymbol{e_\theta})  .
    \label{eqn: simp_body_flow_field_active}
\end{eqnarray}
It is obvious that the Stokeslet term $(\sim r^{-1})$ and Rotlet term $(\sim r^{-2})$ will vanish only if $u_3 = -2\beta/3$ and $\Omega_3=-\gamma/a$ respectively, which corresponds to the velocity and rotation rate components of a fully covered swimmer, as can be obtained by taking limit $(\theta_1,\theta_2) \to (0,\pi)$ on $\boldsymbol{U}$ and $\boldsymbol{\Omega}$ respectively in Eqn.~\ref{eqn: u3_o3}. However, in the present study, the active slip is concentrated over a specific patch region. Consequently, the swimmer’s velocity and rotation rate components are given as Eqn.~\ref{eqn: u3_o3}, and in general, $u_3 \neq -2\beta/3$ and $\Omega_3 \neq -\gamma/a$. Therefore, the Stokeslet $(\sim r^{-1})$ and Rotlet $(\sim r^{-2})$ terms persist in the velocity field $\boldsymbol{u}_\mho$ in the vicinity of the swimmer, despite the swimmer being force-free and torque-free. However, these vanish as $r \to \infty$ in the far field.}

\section{General drag and torque}
\label{sec: Gen_D_T}

In section~\ref{sec: F_T_free_swim}, we briefly describe the analytical calculation for the efficiency of the swimmer that involves the hydrodynamic drag ($\boldsymbol{D_H}$) and torque ($\boldsymbol{T_H}$). For the choice $\beta^A_{11} = 0 =\beta^B_{11}$, $\boldsymbol{D_H}$ and $\boldsymbol{T_H}$, and the drag and torque due to surface activity, $\boldsymbol{D_A}$ and $\boldsymbol{T_A}$, respectively, are given as

\begin{eqnarray}
\boldsymbol{D_H} &=& -6\pi\mu a \bigg(\frac{a+2k}{a+3k}\bigg)\boldsymbol{U}+ \frac{3a\mu k \,s_\phi}{8(a+3k)} \bigg[(\Theta_3+3\Theta_1)\boldsymbol{U} - 3(\Theta_3-\Theta_1){u_3}\boldsymbol{e_3}\bigg]\nonumber\\
&& +\: \frac{3a\mu k (\Theta_3-9\Theta_1)}{16(a+3k)} \bigg[\bigg(\Phi^s_2{u_1} - \Phi^c_2{u_2}\bigg)\boldsymbol{e_1} - \bigg(\Phi^s_2{u_2}+\Phi^c_2{u_1}\bigg)\boldsymbol{e_2}\bigg] + \frac{3a\mu k (\Theta^s_3-3\Theta^s_1)}{4(a+3k)}\nonumber\\
&& \times\: \bigg[-\Phi^s_1{u_3}\boldsymbol{e_1}+\Phi^c_1{u_3}\boldsymbol{e_2}+\bigg(-\Phi^s_1{u_1}+\Phi^c_1{u_2}\bigg)\boldsymbol{e_3}\bigg] + \frac{9a^2\mu k \,s_\phi\, \Theta_2}{4(a+3k)} \bigg[{\Omega_2}\boldsymbol{e_1}-{\Omega_1}\boldsymbol{e_2}\bigg] \nonumber\\
&& -\: \frac{9a^2\mu k}{4(a+3k)}\big(\Theta^s_2-2\,s_\theta\big) \bigg[-\Phi^c_1{\Omega_3}\boldsymbol{e_1}-\Phi^s_1{\Omega_3}\boldsymbol{e_2} + \bigg(\Phi^c_1{\Omega_1} +\Phi^s_1{\Omega_2}\bigg)\boldsymbol{e_3} \bigg],\label{eqn: NAxi_drag_H}\\
\boldsymbol{D_A} &=& \frac{3a\mu\,s_\phi\,\Theta_2}{4}\big(\gamma_{11}^D\boldsymbol{e_1}-\gamma_{11}^C\boldsymbol{e_2}\big) -\frac{3a\mu\,(\Theta^s_2-2\,s_\theta)}{4}  \bigg[\gamma_{10}^C\big(\Phi^c_1\boldsymbol{e_1}+\Phi^s_1\boldsymbol{e_2}\big)
 + \big(\gamma_{11}^C\Phi^c_1+\gamma_{11}^D\Phi^s_1\big)\boldsymbol{e_3}\bigg] \nonumber \\
 && -\: \frac{a\mu}{2}\bigg[ 
 \big(\Theta^s_3-3\Theta^s_1\big)\,\beta_{10}^A\big(\Phi^s_1\boldsymbol{e_1}-\Phi^c_1\boldsymbol{e_2}\big) + s_\phi(\Theta_3-3\Theta_1)\beta_{10}^A\boldsymbol{e_3}\bigg],\label{eqn: NAxi_drag_A}
\end{eqnarray}
\begin{eqnarray}
\boldsymbol{T_H} &=& -\frac{8a^4\pi\mu}{a+3k} \boldsymbol{\Omega} 
- \frac{45a^4\mu k}{4(a+3k)} (\Theta^s_2-2\,s_\theta) \bigg[\Phi^c_1{\Omega_3}\boldsymbol{e_1}+\Phi^s_1{\Omega_3}\boldsymbol{e_2} - \bigg(\Phi^c_1{\Omega_1}+\Phi^s_1{\Omega_2}\bigg)\boldsymbol{e_3}\bigg] \nonumber \\
&& -\: \frac{9a^2\mu k}{8(a+3k)}
\bigg[ s_\phi \Theta_2 ({u_2}\boldsymbol{e_1} - {u_1}\boldsymbol{e_2} ) +
(\Theta^s_2 - 2\,s_\theta) \Big\{\Phi^c_1{u_3}\boldsymbol{e_1}+\Phi^s_1{u_3}\boldsymbol{e_2} - (\Phi^c_1{u_1}+\Phi^s_1{u_2})\boldsymbol{e_3} \Big\} \bigg] \nonumber\\
&& +\: \frac{3a^3\mu k}{16(a+3k)}\bigg[2s_\phi(\Theta_3+15\Theta_1)\boldsymbol{\Omega} - 6s_\phi(\Theta_3-\Theta_1){\Omega_3}\boldsymbol{e_3}
+ (\Theta_3-9\Theta_1)
 \Big\{ (\Phi^s_2{\Omega_1}-\Phi^c_2{\Omega_2})\boldsymbol{e_1} \nonumber\\
 && - (\Phi^s_2{\Omega_2}+\Phi^c_2{\Omega_1})\boldsymbol{e_2} \Big\} 
 - 4(\Theta^s_3-3\Theta^s_1) \Big\{\Phi^s_1{\Omega_3}\boldsymbol{e_1}-\Phi^c_1{\Omega_3}\boldsymbol{e_2} + (\Phi^s_1{\Omega_1}-\Phi^c_1{\Omega_2})\boldsymbol{e_3}\Big\} \bigg],\label{eqn: NAxi_torque_H}
\end{eqnarray}
\begin{eqnarray}
\boldsymbol{T_A} &=& \frac{a^2\mu}{8}
\bigg[ 
s_\phi(\Theta_3+15\Theta_1)(\gamma_{11}^C\boldsymbol{e_1}+\gamma_{11}^D\boldsymbol{e_2}) + 2 s_\phi (\Theta_3-9\Theta_1)\gamma_{10}^C\boldsymbol{e_3}
 - 6\beta_{10}^A(\Theta^s_2-2s_\theta)(\Phi^c_1\boldsymbol{e_1} + \Phi^s_1\boldsymbol{e_2} ) \bigg] \nonumber\\
&& +\: \frac{a^2\mu}{16}\big(\Theta_3-9\Theta_1\big)\bigg(\big(-\gamma_{11}^D\Phi^c_2+\gamma_{11}^C\Phi^s_2\big)\boldsymbol{e_1} - \big(\gamma_{11}^C\Phi^c_2+\gamma_{11}^D\Phi^s_2\big)\boldsymbol{e_2}\bigg) \nonumber\\
&& +\: \frac{a^2\mu}{4}\big(\Theta^s_3-3\Theta^s_1\big)\bigg(\gamma_{10}^C\Phi^s_1\boldsymbol{e_1}-\gamma_{10}^C\Phi^c_1\boldsymbol{e_2}-\big(-\gamma_{11}^D\Phi^c_1+\gamma_{11}^C\Phi^s_1\big)\boldsymbol{e_3}\bigg),
\label{eqn: NAxi_torque_A}
\end{eqnarray}
where $s_\theta = \theta_2 - \theta_1$, $s_\phi = \phi_2 - \phi_1$, 
$\Theta_k = \cos k\theta_2-\cos k\theta_1, \Theta^s_k = \sin k\theta_2-\sin k\theta_1,\Phi^c_k = \cos k\phi_2-\cos k\phi_1,\Phi^s_k = \sin k\phi_2-\sin k\phi_1$ for $k=1,2,3$, and $\left[\boldsymbol{u_\infty}\right]_0 = -\boldsymbol{U} = 2a_1 (-A_{11}\boldsymbol{e_1}-B_{11}\boldsymbol{e_2}+A_{10}\boldsymbol{e_3}), \left[{\boldsymbol{\nabla} \times u_\infty}\right]_0 = -2\boldsymbol{\Omega} = 2e_1 (-C_{11}\boldsymbol{e_1}-D_{11}\boldsymbol{e_2}+C_{10}\boldsymbol{e_3})$. Here $[~\cdot~]_0$ denotes a vector quantity computed at the origin of the body frame of reference attached to the center of the swimmer, which can be expressed as $[~\cdot~]_{0}=[~\cdot~]_{0,{1}}\boldsymbol{e_1}+[~\cdot~]_{0,{2}}\boldsymbol{e_2}+[~\cdot~]_{0,{3}}\boldsymbol{e_3}$, where, $[~\cdot~]_{0,{i}}$ for $i=1,2,3$ are the scalar components of $[~\cdot~]_0$ in the Cartesian coordinates system.





%

\end{document}